\documentclass[aip,rsi,reprint,longbibliography]{revtex4-1}

\usepackage{amsmath}
\usepackage{amsfonts}

\usepackage{microtype}

\usepackage{graphicx}
\usepackage[dvipsnames]{xcolor}

\usepackage{multirow,bigdelim}

\usepackage[urlcolor=blue, colorlinks=true, bookmarks=false]{hyperref}

\newcommand{\ket}[1]{\left\lvert #1 \right\rangle}
\newcommand{\units}[1]{\,\mathrm{#1}}
\newcommand{\psiz}{\psi_0}

\newcommand{\CPP}{C\nobreak\hspace{-.05em}\raisebox{.4ex}{\tiny\bf +}\nobreak\hspace{-.10em}\raisebox{.4ex}{\tiny\bf +}}

\begin{document}

\title{Hardware for Dynamic Quantum Computing}

\author{Colm A. Ryan}
\author{Blake R. Johnson}
\email[Corresponding author: ]{blake.johnson@raytheon.com}
\author{Diego Rist\`e}
\author{Brian Donovan}
\author{Thomas A. Ohki}
\affiliation{Raytheon BBN Technologies, Cambridge, MA 02138, USA}

\date{\today}

\begin{abstract}

We describe the hardware, gateware, and software developed at Raytheon BBN
Technologies for dynamic quantum information processing experiments on
superconducting qubits. In dynamic experiments, real-time qubit state
information is fedback or fedforward within a fraction of the qubits' coherence
time to dynamically change the implemented sequence. The hardware presented here
covers both control and readout of superconducting qubits. For readout we
created a custom signal processing gateware and software stack on commercial
hardware to convert pulses in a heterodyne receiver into qubit state assignments
with minimal latency, alongside data taking capability.  For control, we
developed custom hardware with gateware and software for pulse sequencing and
steering information distribution that is capable of arbitrary control flow on a
fraction superconducting qubit coherence times. Both readout and control
platforms make extensive use of FPGAs to enable tailored qubit control systems
in a reconfigurable fabric suitable for iterative development.

\end{abstract}

\maketitle

\section{Introduction}

Building a large scale quantum information processor is a daunting technology
integration challenge. Most current experiments demonstrate static circuits,
where a pre-compiled sequence of gates is terminated by qubit measurements. In
some cases, conditional control flow is emulated by postselecting data on
certain measurement outcomes~\cite{Chow14}, or by gating duplicate hardware
behind a switch to handle a single branch in a pulse
program~\cite{Steffen:2013,Riste2013}. However, because of the need for quantum
error correction~\cite{Gottesman2009}, fault-tolerant quantum computation is
inevitably an actively controlled process. This active control may manifest as:
continuous entropy removal from the system via active reset \cite{Aharonov1996},
active error correction after decoding syndrome measurements, Pauli frame
updates for subsequent pulses after state injection~\cite{Bravyi2005,
Knill2005}, or non-deterministic ``repeat-until-success''~\cite{Paetznick:2014}
gates. The community is now tackling the challenge of dynamically steering an
experiment within the coherence time of the
qubits~\cite{Riste2012,CampagneIbarcq2013,Pfaff2014,Ofek2016}. For
superconducting qubits this coherence time---although continuously
improving---is currently 50--100$\units{\mu s}$. To achieve control fidelities
compatible with expected thresholds for fault-tolerant quantum
computation~\cite{Knill2005,Raussendorf2007b}, the feedback/feedforward time
must be less than 1\% of this coherence time, or on the order of a few hundred
nanoseconds.

Superconducting qubit control systems send a coordinated sequence of microwave
pulses, with durations from tens to hundreds of nanoseconds, down coaxial lines
of a dilution refrigerator to implement both control and readout of the qubits.
Currently, the microwave pulses are produced and recorded at r.f. frequencies by
mixing up or down with a microwave carrier, allowing commonly available
$\approx1$ GS/s digital-to-analog (DAC) and analog-to-digital (ADC) converters
to be used. In the circuit quantum electrodynamics (QED)
platform~\cite{Schuster2007}, the qubit state is encoded in the amplitude and
phase of a measurement pulse that interacts with a microwave cavity coupled
dispersively to the qubit. This microwave pulse is typically captured with a
room temperature receiver, then converted into a qubit state assignment via a
digital signal processing (DSP) pipeline. Programming the control sequences for
dynamic experiments also requires a supporting framework from the pulse
sequencing language and hardware. Conventional arbitrary waveform generator
(AWG) sequence tables are far too restrictive to support control flow beyond
simple repeated sections. The desired control flow requires conditional
execution, loops with arbitrary nesting, and subroutines for code reuse.

The required timescale for active control is beyond the capabilities of a
software solution running on a general purpose operating system; however, it is
within reach of custom gateware running on field programmable gate arrays
(FPGAs) directly connected to analog $\leftrightarrow$ digital converters for
both qubit control and measurement. Many groups in superconducting and ion trap
quantum computing have turned to this approach and started to build a framework
of controllers and actuators. For trapped ions, the Advanced Real-Time
Infrastructure for Quantum physics (ARTIQ)~\cite{artiq} is a complete framework
of hardware, gateware, and software for controlling quantum information
experiments. However, ARTIQ's control flow architecture uses general purpose
CPUs implemented in FPGA fabric, so called \emph{soft-core} CPUs, which cannot
maintain the event rate required by superconducting qubits (gates are 1--2
orders of magnitude slower in ion traps). Researchers at
UCSB/Google~\cite{Chen:2012,Jeffrey:2014}, ETH Zurich~\cite{Steffen:2013}, TU
Delft~\cite{Riste2013,Bultink:2016}, and Yale~\cite{Ofek2016} have also built
superconducting qubit control and/or readout platforms using FPGAs, and even
explored moving them to the cryogenic stages~\cite{Homulle:2016,ConwayLamb2016},
but have generally not made these tools available to the broader quantum
information community.

\begin{figure}
\includegraphics[width=\columnwidth]{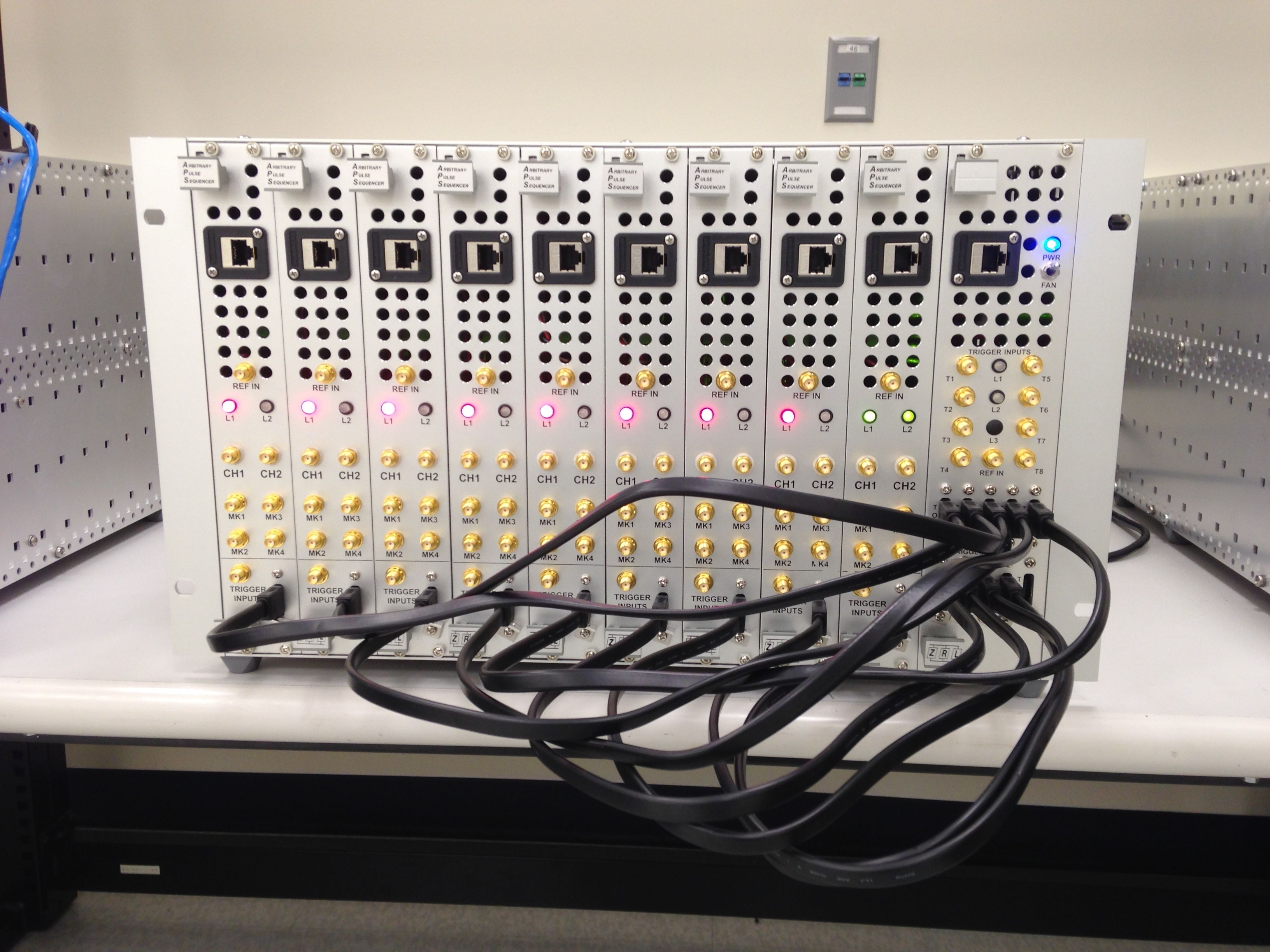}
\caption{\label{fig:APS2} A fully-populated APS2 system, with 18 analog output
channels (9 APS2 modules) and a trigger distribution module (TDM, far right).
Each APS2 module provides two 14-bit analog output channels with 1.2 GS/s
sampling rate and four digital marker channels. The 6U enclosure provides power
and cooling. Inter-module signaling is handled by the star network of SATA
cables between the TDM and each APS2 output module. Host control is via 1Gb
Ethernet to each module with a combination of a Comblock 5402 TCP stack,
\texttt{APS2-Comms} custom HDL (\url{github.com/BBN-Q/APS2-Comms}) and the
\texttt{libaps2} \protect\CPP~software driver (\url{github.com/BBN-Q/libAPS2})}
\end{figure}

In this work, we introduce the \texttt{QDSP} framework and Arbitrary Pulse
Sequencer 2 (APS2) for qubit readout and control, respectively. \texttt{QDSP}
implements state assignment and data recording in FPGA gateware for a
commercially available receiver/exciter system (the Innovative Integration
X6-1000M, also used in the Yale work\cite{Ofek2016}). We show how latency can be
minimized for rapid qubit state decisions by consolidating many of the
conventional DSP stages into one. The APS2, shown in Fig.~\ref{fig:APS2}, has
gateware designed to naturally support arbitrary control flow in quantum circuit
sequences on superconducting qubits. For circuits involving multiple qubits,
state information from many qubits must be collated and synthesized into a
steering decision by a controller. To this end we designed the Trigger
Distribution Module (TDM) to capture up to eight channels of qubit state
information, execute arbitrary logic on an FPGA, and then distribute steering
information to APS2 output modules over low-latency serial data links. All the
systems presented here are either commercially available or full source code for
gateware and drivers has been posted under a permissive open-source license.

To validate the developed gateware and hardware we demonstrate multi-qubit
routines and quantum gates that require feedback and feedforward: active qubit
initialization, entanglement generation through measurement, and
measurement-based logic gates. Although these are specific examples, they are
implemented in a general framework that enables arbitrary steering of quantum
circuits. Furthermore, with appropriate quantum hardware, different circuits
are all achieved without re-wiring the control systems, but simply by executing
different programs on the APS2 and TDM.

\section{Qubit State Decisions in Hardware}
\label{sec:qubit-state-decision}

The first requirement for quantum feedback is extracting qubit state decisions
with minimal latency. Typical superconducting qubit measurements involve sending
a microwave pulse to a readout resonator, recording the reflected/transmitted
signal, filtering noise and other out-of-band signals, and reducing the record
to a binary decision about the qubit state. Conventionally, this is accomplished
with a superheterodyne transmitter and receiver operating with an intermediate
frequency (IF) of 10s of MHz which allows the IF stages to be handled
digitally.

\begin{figure}
\includegraphics[width=\columnwidth]{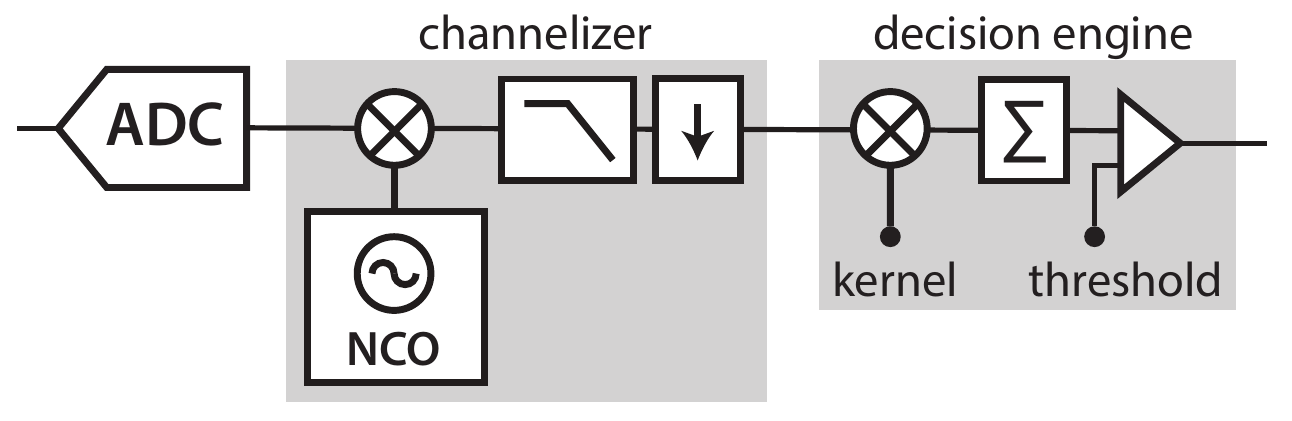}
\caption{ \label{fig:qubit-dsp} Block diagram of typical superheterodyne
receiver qubit decision chain: filtering/decimation, demodulation,
filtering/decimation, integration and thresholding. The filtering and decimating
may be combined into a polyphase decimating filter. This filter may also consist
of multiple stages for stability or efficiency reasons.}
\end{figure}

Since many measurement channels may be frequency multiplexed onto the same line,
the DSP chain involves several stages of filtering to \emph{channelize} the
signal. This involves mixing the captured record with a continuous wave (CW) IF
signal---produced by a numerically controlled oscillator (NCO)---and several
low-pass filtering and decimation stages to recover a baseband complex-valued
phasor as a function of time (Fig.~\ref{fig:qubit-dsp}). This complex-time
series is then integrated with a kernel, which may be a simple box car filter or
optimized to maximally distinguish the qubit
states~\cite{Gambetta2007,Ryan2015}. A final qubit state is determined by
thresholding the integrated value. These receiver functions, which have
frequently been implemented in software, are ideally suited to DSP resources
available in modern FPGAs. Moving these functions into custom gateware has
additional benefits for parallel processing of simultaneous measurements,
reducing CPU load on the control PC, and greatly reducing latency of qubit state
decisions.

\subsection{Filter design}

The design of the channel filter for qubit readout is the result of balancing
several considerations:
\begin{enumerate}
\item bandwidth of the channel---should be some small multiple of the resonator
  bandwidth, $\kappa$;
\item stopband attenuation sufficient to remove channel crosstalk;
\item numerical stability---particularly when implemented with either single
  precision or fixed-point representation;
\item latency;
\item computational resources.
\end{enumerate}

Some of these criteria are in competition with each other. For instance, one may
decrease channel crosstalk by using a higher-order filter, but this comes at the
expense of increased latency and computational cost. Qubit devices used in our
lab have typical resonator bandwidths of $1-3\units{MHz}$. In the high fidelity,
QND readout regime we have noticed harmonic content in the readout signal at
multiples of the dispersive shift, $\chi$, that extends the signal bandwidth by
roughly a factor of 2. Consequently, we have designed channel filters with
$10\units{MHz}$ bandwidth. The downconversion structure of
Fig.~\ref{fig:qubit-dsp} selects symmetric channels around the IF frequency;
thus, a $10\units{MHz}$ channel corresponds to a filter with a $3\units{dB}$
bandwidth of $5\units{MHz}$. We also want sufficient stopband attenuation to
limit channel crosstalk. We have chosen the stopband attenuation such that a
fullscale signal in an adjacent channel is suppressed below the
least-significant bit of the selected channel. Given the signed 12-bit ADCs on
our target platform, this requires $20\log_{10}(1/2^{11}) \approx 66\units{dB}$
stopband attenuation.

The relatively narrow bandwidth of the readout channels compared to the 1 GS/s
sampling rate of the ADC leads to numerical stability problems in fixed-point or
single-precision designs. Re-expressed as a relative bandwidth, the
$f_{3\mathrm{dB}} = 5$ MHz channel described above has $n_{3\mathrm{dB}} =
0.01$. However, it is difficult to construct stable filters with normalized
bandwidth $n_{3\mathrm{dB}}<0.1$. This may be solved by cascading several
polyphase decimating filters to boost the 3 dB bandwidth of the later stages---
this brings an additional benefit of reducing the computational resources.

\subsection{Fast Integration Kernels}

While the complete time-trace of the measurement record is a useful debugging
tool for observing and understanding the cavity response from the two (or more)
qubit states, a conventional channelizer with multiple stages of signal
processing (NCO mixing, filtering and integrating) forces an undesirable
latency. Take a typical example of $10\units{MHz}$ channels spaced
$20\units{MHz}$ apart. A Parks-McClellan \cite{McClellan1973} designed FIR
low-pass filter for a $250\units{MHz}$ sampling rate with a pass band from
$0-5\units{MHz}$ and stop-band from $15-125\units{MHz}$ with $60\units{dB}$
suppression requires at least 86 taps. At a typical FPGA clock speed of
$250\units{MHz}$ this results in 100s of nanoseconds of latency. However, the
qubit state decision reduces the time dimension to a single value with a kernel
integrator. The intermediate filtering stage is thus superfluous if we can
construct an appropriate frequency-selective kernel. This crucial insight
enables us to drive down the signal processing latency to a few clock cycles.


More formally, consider the discrete time measurement record $v(t_l)$ for a
total of length $L$ samples. Applying the DSP chain of Fig.~\ref{fig:qubit-dsp} ,
the final single complex value qubit state signal (before thresholding and
ignoring decimation for simplicity) is:
\begin{equation}
\label{eq:3stage-dsp}
q = \underbrace{\sum_{l=0}^L k_l}_{\text{kernel}} \left[\underbrace{\sum_{n=0}^N b_n }_{\text{filter}} \left[ \underbrace{e^{-i\omega (t_{l-n})}}_{\text{mix-down}} v(t_{l-n}) \right]\right],
\end{equation}
where the demodulation frequency is $\omega$, the channel is selected with an
$N$-tap FIR filter with coefficients $b_n$ and a final kernel integration $k_l$
is applied for the length of the record $L$. The nested sum and product can be
expanded and the terms collected into a single kernel integration, with a modified kernel
\begin{equation}
\label{eq:modified-kernel}
q = \sum_{l=0}^{L} k_l' v(t_l); \quad k_l' = e^{-i\omega t_l}\sum_{n=0}^N k_{l+n} b_n.
\end{equation}
Thus, the three-stage pipeline of Fig.~\ref{fig:qubit-dsp} is reduced
into a single-stage pipeline consisting solely of the kernel integration
step.

This reduction of the pipeline to a single stage has substantial advantages for
DSP latency. In particular, the FIR filter block of the three-stage pipeline has
a minimum latency of N clock cycles for a N-tap filter. As discussed above, this
can be 100s of nanoseconds and this single filter stage consumes the entire
latency budget in a single step. By constrast, the DSP pipeline of
Eq.~\ref{eq:modified-kernel} can be achieved with 1-3 clock cycles of latency on
the FPGA, or $\le 15\units{ns}$.

While equations \ref{eq:3stage-dsp} and \ref{eq:modified-kernel} demonstrate the
mathematical equivalence of the 1-stage and 3-stage DSP pipelines, in practice
it is not necessary to transform a baseband integration kernel via
Eq.~\ref{eq:modified-kernel}. Instead, one can use the average unfiltered (IF)
response at the ADC after preparing a qubit in $\ket{0}$ and $\ket{1}$ to
construct a matched filter \cite{Ryan2015}. The frequency response of the
resulting filter will match that of the measurement pulse itself. Consequently,
as long as the measurement pulse is itself band-limited --- which should always be
the case with an appropriately designed dispersive cavity measurement --- the
resulting matched filter will also optimally ``channelize'' the ADC input and
suppress interference from other multiplexed qubit measurement channels.

\subsection{Hardware Implementation}

\begin{figure}
\includegraphics[width=\columnwidth]{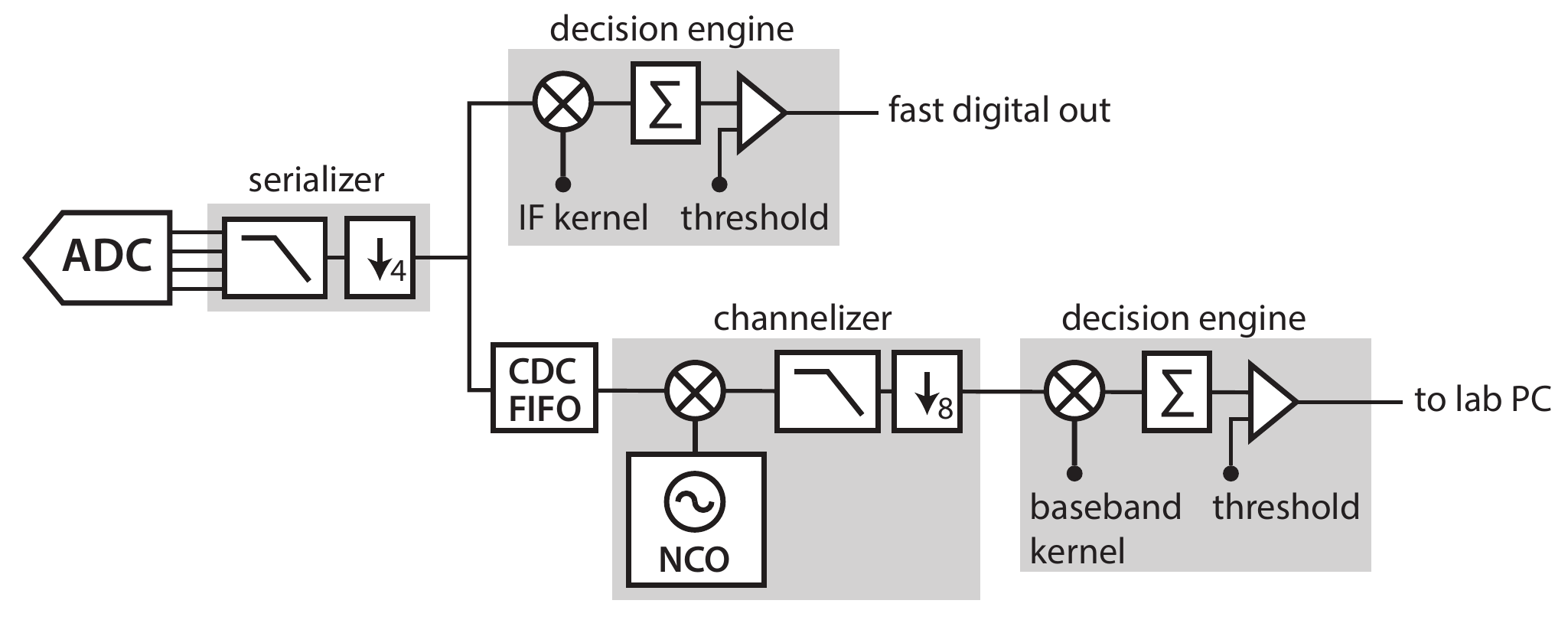}
\caption{ \label{fig:X6-block-diagram} Block diagram of \texttt{QDSP} filter
blocks, with low-latency feedback (upper) and calibration/diagnostic (lower)
paths. The low-latency path drives digital outputs which may be connected to the
control system, such as the TDM (see section~\ref{sec:TDM}). $N$ copies (not
shown) of this low-latency path support multiplexed readout. The clock-domain
crossing (CDC) FIFO on the diagnostic path allows the low-pass filter in the
channelizer to run at a slower clock rate, easing timing closure. Both slow and
fast paths are duplicated four times per ADC in order to handle multiplexed
signals. The user may choose to tap these data streams at various points, and
send the data over PCIe for recording on the host PC.} \end{figure}

To minimize overall latency, we implement our \texttt{QDSP} qubit readout system in
custom FPGA gateware (\texttt{QDSP} - \url{github.com/BBN-Q/BBN-QDSP-X6}) and
software drivers (\texttt{libx6} - \url{github.com/BBN-Q/libx6/}) for a
commercially available hardware platform (Innovative Integration X6-1000M). The
X6 hardware provides two 12-bit 1 GS/s ADCs and four 16-bit 500 MS/s DACs.
Although \texttt{QDSP} focuses on the receiver
application, it also provides basic AWG functionality to drive the DACs for
simple waveforms such as measurement pulses. A block diagram of the receiver
section of the \texttt{QDSP} gateware is shown in Fig.~\ref{fig:X6-block-diagram}. The
structure includes a fast path for low-latency qubit state decision output, as
well as a conventional receiver chain for debugging and calibration. The
gateware and drivers allow users to tap the data stream at several points for
data recording or debugging.

The raw ADC values from each ADC are presented to the FPGA four samples wide at
250 MHz when sampling at 1 GS/s (we sample at the maximum rate to minimize noise
aliasing).  We immediately decimate by a factor of 4 by summing the four values
so that subsequent processing deals with only one sample per clock. This is
mainly for convenience: the raw integrators could run in parallel and the data
could be serialized for the subsequent filtering. The data is copied to $N$ IF
kernel integrators for multiplexed readout. The outputs of these fast
integrators are connected to variable thresholders which drive digital outs to
make fast qubit state decisions available to the pulse sequencing hardware for
feedback. These values are also available in software as complex values.

For more conventional downconversion, each raw stream is also broadcast to a
channelizer module. The module consists of a numerically controlled oscillator
(NCO) that generates cosine and sine at the chosen frequency.  The incoming ADC
data is multiplied with the NCO outputs in a complex multiplier. The mixed
signal is then low-pass filtered by a two-stage decimating finite-impulse
response (FIR) filter chain. Polyphase FIR filters are chosen for each stage to
minimize use of specialized DSP hardware on the FPGA. The FIR filters are
equiripple with the coefficients designed by the Remez exchange algorithm
\cite{McClellan1973}. The number of taps was chosen to optimally fit onto the
DSP blocks of the FPGA (with reuse from hardware oversampling) and to suppress
the stopband by 60 dB, nearly down to the bit level of the 12-bit ADCs. The
low-pass filtered and decimated stream is useful for observing and debugging the
cavity response. Finally, a decision engine using a baseband kernel integrator
is attached to the demodulated stream to complete the conventional DSP chain.

\section{Dynamic Arbitrary Pulse Sequencing: APS2}

There are demanding requirements on bandwidth, latency and noise for dynamic
pulse sequencing with superconducting qubits. The sequencer should naturally
represent the quantum circuit being applied, i.e., it should be able to apply a
sequence of $\approx 20\units{ns}$ pulses (typical single qubit gate times)
rather than treating the entire sequence as a waveform. Simply concatenating
waveforms together to create a sequence places extreme demands on the size of
waveform memory, and transferring and compiling sequences to the AWG becomes an
experimental bottleneck. The sequencer should be able to respond to real-time
information from qubit measurement results to make dynamic sequence selection
within some small fraction of the relaxation time of the qubits. Finally, the
sequencer output should have sufficiently low noise not to limit gate fidelity.

Typical AWGs rely on a precalculated list of sequences played out in a
predetermined manner, or at best, loops of segments with simple jump responses
to an event trigger. Dynamic sequences that implement quantum algorithms require
more sophisticated control flow with conditional logic and branching in response
to measurement results. In addition to dynamic control flow, the sequencer
should also support code reuse through function calls and looping constructs to
keep memory requirements reasonable for long verification and validation
experiments such as randomized benchmarking~\cite{Knill2008} or gate set
tomography~\cite{Blume-Kohout2013}.

\begin{figure}
\includegraphics[width=\columnwidth]{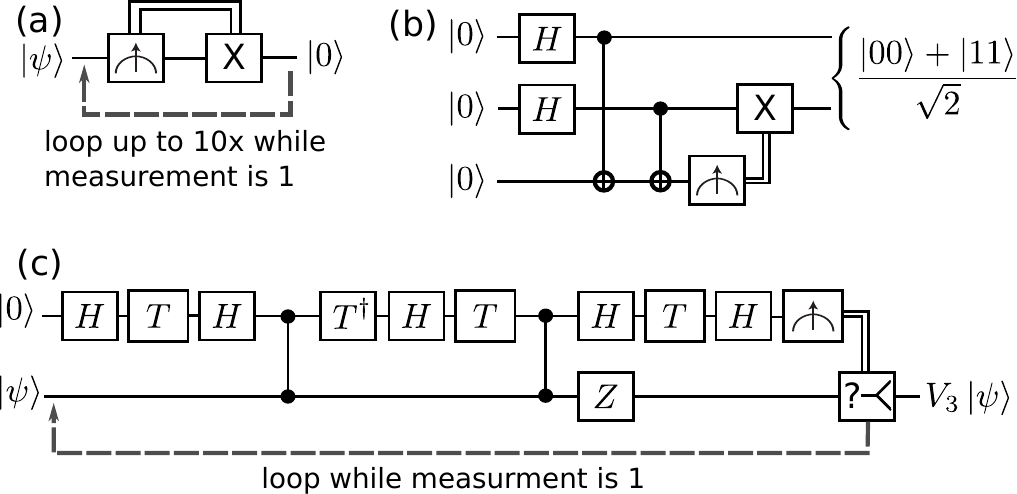}
\caption{ \label{fig:dynamic-quantum-circuits} Example circuits with dynamic
steering: (a) active qubit reset; (b) deterministic entanglement creation with
feedforward; (c) ``repeat-until-success'' implementation of a non-Clifford gate
$V_3 = \frac{1+2iZ}{\sqrt{5}}$.}
\end{figure}

Figure~\ref{fig:dynamic-quantum-circuits} shows some elementary circuits that
require fast feedback or feedforward. A simple and immediately useful primitive
is the active reset of a qubit shown in
Fig.~\ref{fig:dynamic-quantum-circuits}(a). This can remove entropy from the
system by refreshing ancilla qubits or simply improve the duty cycle of an
experiment in comparison to waiting several multiples of $T_1$ for the qubit to
relax to the ground state. With appropriate control flow instructions, reset
with a maximum number of tries is naturally expressed as a looping construct
with conditional branching for breaking out of the loop. Indeed the entire
routine could be wrapped as a function call to be reused at the beginning of
every sequence. Entanglement generation by measurement, shown in
Fig.~\ref{fig:dynamic-quantum-circuits}(b) is another useful primitive for
resource state production that relies on feedforward. The circuit is also a
useful testbench as it is very similar to the circuits for syndrome measurement
in error correcting codes. Finally, Fig.~\ref{fig:dynamic-quantum-circuits}(c)
shows a more sophisticated use of feedforward. Implementing T gates will most
likely dominate the run time of an error corrected quantum circuit
\cite{Raussendorf2007a}. However, if the circuit can be probabilistic then the
average T gate depth can be reduced. These ``repeat-until-success''
circuits~\cite{Paetznick:2014} bring in one or more ancilla qubits and perform a
series of gates and interactions. Then, \emph{conditional} on the result of
measuring the ancilla either the desired gate or a identity operation has been
applied to the data qubit. In the identity case, the gate can be attempted again
by repeating the circuit with a refreshed ancilla.

The APS2 was constructed to satisfy all these criteria by tailored design of the
sequencer. The sequencing engine processes an instruction set that provides full
arbitrary control flow and can play a new waveform every $6.66\units{ns}$ (two FPGA
clock cycles) to naturally and compactly represent any superconducting qubit
circuit with feedback or feedfoward. Realtime state information is fed in via
high-speed serial links from the TDM. A cache controller intermediates access to
deep memory for longer experiments. We now discuss in detail some of the design
choices.

\subsection{Arbitrary Control Flow}
\label{sub:arbitrary-control-flow}

Arbitrary control flow can be fulfilled with three concepts: sequences, loops
(repetition) and conditional execution. We add to this set the concept of
subroutines because of their value in structured programming and memory re-use.
The gateware implements a control unit state machine with four additional
resources: a loadable incrementing program counter indicating the current
address in instruction memory; a loadable decrementing repeat counter; a stack
that holds the repeat and program counter values; and a comparison register that
holds the last comparison boolean result. The specific instruction set supported
is shown in Table~\ref{table:APS2-instruction-set}.

\begin{table*}
  \begin{ruledtabular}
    \begin{tabular}{r p{1.5\columnwidth} l}
      \texttt{WAVEFORM} & dispatch instruction to waveform engine(s) & \rdelim\}{3}{25mm}[output] \\
      \texttt{MARKER} & dispatch instruction to marker engine(s) \\
      \texttt{MODULATOR} & dispatch instruction to I/Q modulator engine \\
      \texttt{WAIT} & broadcast wait command to all output engines & \rdelim\}{2}{25mm}[synchronization]  \\
      \texttt{SYNC} &  wait until all execution engine queues are empty \\
      \texttt{LOAD\_REPEAT} & load value into the repeat register & \rdelim\}{9}{25mm}[flow control] \\
      \texttt{REPEAT} & if repeat register is 0 continue; otherwise decrement repeat register and jump to given address  \\
      \texttt{LOAD\_CMP} & load comparison register with next value in serial link FIFO \\
      \texttt{CMP} & compare given mask to comparison register with given binary comparison operation ($ =$, $\ne$, $<$, $>$) and store result in  comparison result register \\
      \texttt{GOTO} & jump to given instruction address (optionally conditionally) \\
      \texttt{CALL} & push stack and jump to the given instruction address (optionally conditionally) \\
      \texttt{RETURN} &  pop stack and return to the call site \\
      \texttt{PREFETCH} & prefetch an instruction cache line
    \end{tabular}
  \end{ruledtabular}
  \caption{The APS2 instruction set which enables arbitrary control flow with waveform generation.}
  \label{table:APS2-instruction-set}
\end{table*}

The \texttt{WAVEFORM}, \texttt{MARKER} and \texttt{MODULATOR} instructions
enable analog and digital output and are immediately dispatched to output
execution engines (see sections~\ref{subs:superscalar-architecture} and
\ref{subs:output-engines} below). The next two instructions, \texttt{WAIT} and
\texttt{SYNC}, enable synchronization both between output engines on the same
APS2 and between APS2 modules (see section~\ref{subs:superscalar-architecture}
below). The next set of instructions provides arbitrary control-flow:
\texttt{LOAD\_REPEAT} and \texttt{REPEAT} enable looping constructs;
\texttt{LOAD\_CMP} enables access to the real-time steering information fed from
the TDM; \texttt{CMP} and \texttt{GOTO} enable conditional branching;
\texttt{CALL} and \texttt{RETURN} allow for subroutines and recursion, enabling,
for example, nested loops without multiple loop counters. Finally, although not
directly related to control flow, \texttt{PREFETCH} gives hints to the cache
controller to avoid cache misses.

\subsubsection{Super-scalar Architecture}
\label{subs:superscalar-architecture}

\begin{figure}
\includegraphics[width=0.95\columnwidth]{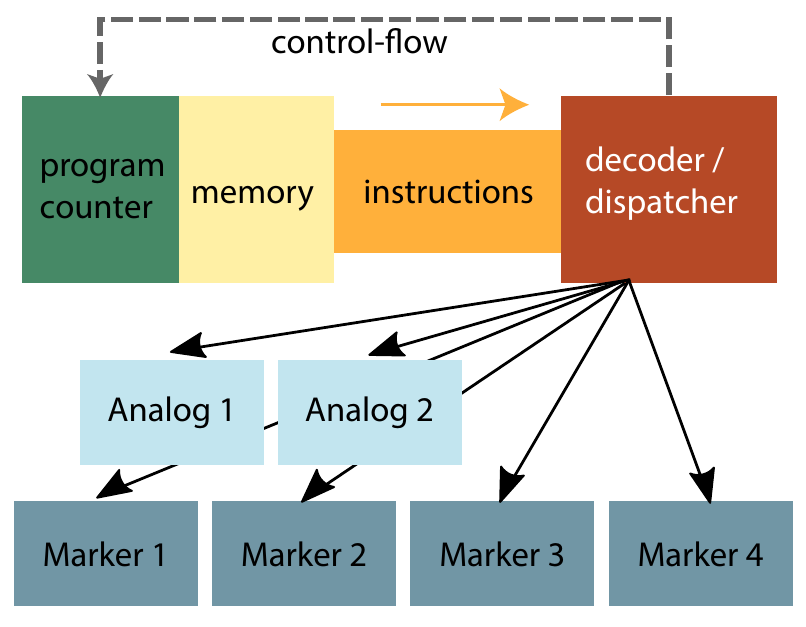}
\caption{ \label{fig:superscalar-architecture} The APS2 has a superscalar
architecture where a linear instruction stream is dispatched to multiple
execution engines which then execute in parallel. The program counter increments
by default sending a stream of instruction to the decoder/dispatcher.
Control-flow instructions can cause the decoder to jump the program counter and
flush the instruction stream coming from memory.}
\end{figure}

Each APS2 module has multiple outputs driven by individual execution engines:
two analog channels and four marker channels. We use dispatch from a single
instruction stream to simplify synchronization of control flow across multiple
output engines (Fig.~\ref{fig:superscalar-architecture}). Since each execution
engine has its own internal FIFO buffer, this also allows the decoder/dispatcher
to greedily look ahead and process instructions (contingent on deterministic
control flow) and potentially dispatch to the execution units. The look ahead
strategy absorbs the pipelining latency due to an instruction counter address
jump after a \texttt{CALL}, \texttt{RETURN} or \texttt{REPEAT} instruction.

The superscalar approach has to accept some additional complexity in order to
convert a serial instruction stream into potentially simultaneous operations in
the execution engines. The APS2 provides two mechanisms to solve this
synchronization task. The first mechanism is a \texttt{WAIT} instruction that
stalls the execution engines until a trigger signal arrives. While the engines
are stalled, the control flow unit/dispatcher continues to load instructions
into the output engine buffers. The execution engines respond synchronously to
trigger signals, so in this mechanism an external signal provides simultaneity
and a method to synchronize multiple modules. The second mechanism, the
\texttt{SYNC} instruction, acts as a fence or barrier to ensure that all
execution engines are at the same point by stalling processing of instructions
until all engines' execution queues are empty. This is also useful for
resynchronizing after a non-deterministic wait time - e.g. an uncertain delay
before a measurement result is valid.

\subsubsection{Output Engines}
\label{subs:output-engines}

Each analog and digital output channel is sequenced by a waveform or marker
``output engine'' that takes a more limited set of instructions.

\paragraph{Waveform Engine}
\label{par:Waveform Engine}

The waveform engines create analog waveforms from the following set of instructions:

\begin{enumerate}
  \item \texttt{PLAY} play a waveform starting at a given address for a given count;
  \item \texttt{WAIT} stall playback until a trigger arrives;
  \item \texttt{SYNC} stall until the main decoder indicates all engines are synchronized;
  \item \texttt{PREFETCH} fill a page of the waveform cache from deep memory - see section \ref{subs:waveform-cache} for further details.
\end{enumerate}

Typically, each \texttt{PLAY} instruction corresponds to a pulse implementing a
gate and so it is important that the waveform engine be fed and be able to
process instructions on a timescale commensurate with superconducting qubit
control pulses. The main decoder can dispatch a waveform instruction every
$3.33\units{ns}$ and the waveform engine can jump to a new pulse every
$6.66\units{ns}$. In addition, typical pulse sequences contain idle periods of
zero or constant output. Rather than inefficiently storing repeated values in
waveform memory. Rather the instruction is ``play this waveform value for $n$
samples'' \cite{Morgan1987}. We refer to these as \texttt{Time-Amplitude (TA)}
pairs and can mark any waveform command as such.

\paragraph{Marker Engine}

Marker engines creates digital outputs from the following set of instructions:

\begin{enumerate}
  \item \texttt{PLAY} play marker with a given state for a given count;
  \item \texttt{WAIT} stall playback until a trigger arrives;
  \item \texttt{SYNC} stall until the main decoder indicates all engines are synchronized.
\end{enumerate}

The natural sample rate for the marker \texttt{PLAY} commands are in terms of
the sequencer FPGA clock which runs at a quarter of the analog output rate. To
provide single sample resolution we route the marker outputs through dedicated
serializer hardware (Xilinx OSERDESE2). For all but the last sample the 4 marker
samples are simply copies of the desired output state. However, the last word is
programmable as part of the \texttt{PLAY} instruction to provide full
$833\units{ps}$ resolution of the marker rising/falling edge.

\subsubsection{Modulation Engine}
\label{subs:modulation-engine}

An APS2 module is typically used to drive the I and Q ports of an I/Q mixer to
modulate the amplitude and phase of a microwave carrier, thus producing the
control or readout signal. To improve the on/off ratio, the carrier is typically
detuned from the qubit or cavity frequency and the I/Q waveforms modulated at
the difference frequency with an appropriate phase shift to single-sideband
(SSB) modulate the carrier up or down to the qubit/cavity frequency. Qubit
control is defined in a rotating frame at the qubit frequency so the phase of
the modulation has to track the detuning frequency. Z-rotations are implemented
as frame updates that shift the phase of all subsequent pulses \cite{Knill2000}.
For deterministic sequences, the modulation and frame changes can be
pre-calculated and stored as new waveforms in the pulse library. However, for
conditional execution or for experiments with non-deterministic delays, this is
not possible and the modulation and frame changes must be done in real-time.

To support both SSB modulation and dynamic frame updates, the APS2 includes a
modulation engine which phase modulates the waveform output, and that can be
controlled via sequence instructions. The modulation engine contains multiple
NCOs to enable merging multiple ``logical'' channels at different frequencies
onto the same physical channel pair. For example, to control two qubits, two
NCOs can be set to the detuning frequencies of each qubit, and control pulses
can be sent to either qubit with the appropriate NCO selection, while the
hardware tracks the other qubit's phase evolution. The phase applied to each
pulse is the sum of the accumulated phase increment (for frequency detuning), a
fixed phase offset (e.g. for setting an $X$ or $Y$ pulse), and an accumulated
frame (to implement $Z$-rotations). The modulation engine supports the following
instructions

\begin{enumerate}
  \item \texttt{WAIT} stall until a trigger is received;
  \item \texttt{SYNC} stall until the main decoder indicates all engines are synchronized;
  \item \texttt{RESET\_PHASE} reset the phase and frame of the selected NCO(s);
  \item \texttt{SET\_PHASE\_OFFSET} set the phase offset of the selected NCO(s);
  \item \texttt{SET\_PHASE\_INCREMENT} set the phase increment of the selected NCO(s);
  \item \texttt{UPDATE\_FRAME} update the frame of the selected NCO(s);
  \item \texttt{MODULATE} select a NCO for a given number of samples.
\end{enumerate}

All NCO phase commands are held until the the next instruction boundary, which
is the end of the currently playing \texttt{MODULATE} command or a
synchronization signal being received. The commands are held to allow them to
occur with effectively no delay: for example, the phase should be reset when the
trigger arrives; or a $Z$ rotation should happen instantaneously between two
pulses.

\begin{figure}
\includegraphics[width=0.95\columnwidth]{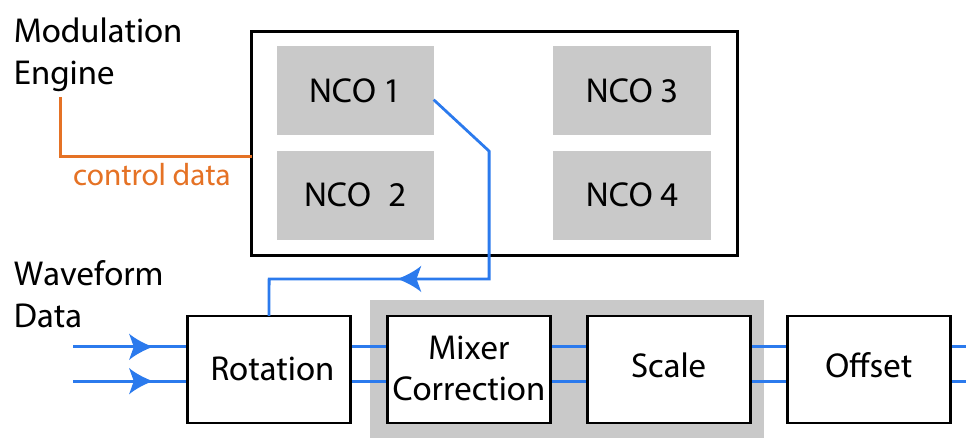}
\caption{ \label{fig:modulator} Block diagram of the APS2 modulation
capabilities. The modulation engine controls the NCO phase accumulators and
selects the desired NCO on a pulse-by-pulse basis. The complex waveform data is
rotated by the selected NCO's phase and subsequently processed by an arbitrary
2x2 matrix for amplitude and phase imbalance correction, channel scaling, and
offset. To save FPGA resources and reduce latency, the scaling is combined with
the mixer correction.}
\end{figure}

In addition, I/Q mixers have imperfections that can be compensated for by
appropriate adjustments to the waveforms. In particular, carrier leakage may be
minimized by adjusting DC offsets, and amplitude/phase imbalance compensated with a
2x2 correction matrix applied to the I/Q pairs. The APS2 includes correction
matrix and offset blocks after the modulator to effect these adjustments, as
shown in Fig.~\ref{fig:modulator}.

\subsection{Caching Strategies}
\label{sub:caching-strategies}

Some qubit experiments, e.g. calibration and characterization, require long
sequences and/or many waveform variants. Supporting such sequences requires an
AWG with deep memory. However, AWG sequencers immediately run into a well-known
depth/speed trade-off for memory: SDRAM with many gigabytes of memory has random
access times of 100s of nanoseconds whereas SRAM, or on-board FPGA block RAM,
can have access times of only a few clock cycles but are typically limited to
only a few megabytes. This memory dichotomy drives some of the sequencing
characteristics of commercial AWGs. For example, the Tektronix 5014B requires
$400\units{ns}$ to switch sequence segments and the Keysight M8190A requires a
minimum sequence segment length of $1.37\units{ms}$. These delay times are
incompatible with the typical gate times of 10s of nanoseconds for
superconducting qubits. However, it is possible to borrow from CPU design and
hide this latency by adding instruction and waveform caches to the memory
interface.

The APS2 has $1\units{GB}$ of DDR3 SDRAM to dynamically allocate to a
combination of sequence instructions and waveforms. This corresponds to up to
128 million sequence instructions or 256 million complex waveform points,
sufficient for most current experiments. The sequencer and waveform engines
interface with this deep memory through a cache controller with access to FPGA
block RAM. If the requested data is in the cache, then it can be returned
deterministically within a few clock cycles, whereas if there is a cache miss
the sequencer stalls while the data is fetched from SDRAM. Cache misses during a
sequence are generally catastrophic given superconducting qubit coherence times.
However, with heuristics and \texttt{PREFETCH} hints from the compiler, the
cache controller can ensure data has been preloaded into the block RAM before it
is requested and avoid any cache stalls.

\subsubsection{Instruction Cache}
\label{subs:instruction-cache}

\begin{figure*}
\includegraphics[width=0.95\textwidth]{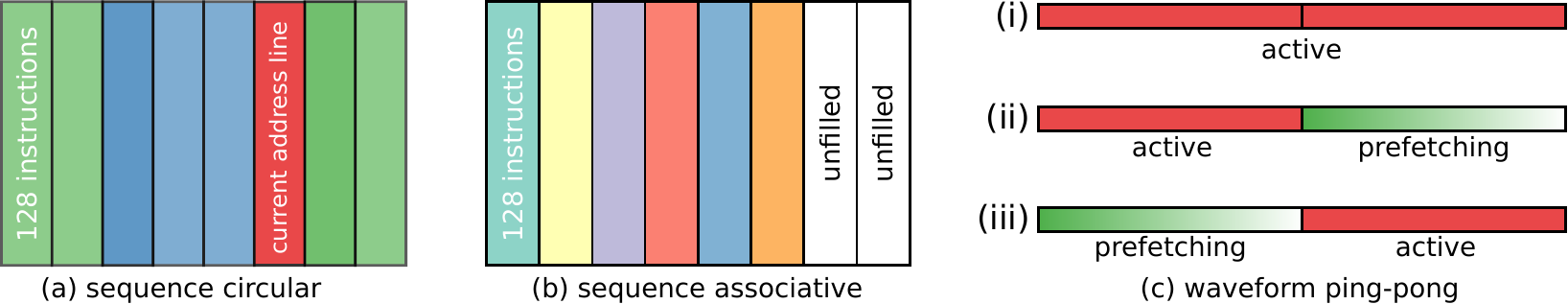}
\caption{ \label{fig:cache-architectures} Instruction and waveform cache
architectures. The instruction cache has two parts: (a) a circular sequential
instruction cache that supports continuous playback by prefetching cache lines
(green) following the cache line containing the currently playing address (red)
up to a local jump buffer of previously played lines (blue); (b) a fully
associative subroutine cache that supports jumps to arbitrary addresses and is
explicitly filled by \texttt{PREFETCH} instructions. (c) The waveform cache
supports either single usage of the full 128 ksamples or a ping-pong mode where
while one half is active the other half is filled by a waveform engine
\texttt{PREFETCH} command.}
\end{figure*}

The APS2 instruction cache is split into two parts to support two different
heuristics about how sequences advance through the instruction stream---see
Fig.~\ref{fig:cache-architectures}(a-b). We chose cache line sizes of 128
instructions or 1\,kB, which is significantly larger than those used in a typical
CPU (Intel/AMD processors typically have cache lines of only 64 bytes) but
reflects the lack of a nested cache hierarchy and the more typical linear
playback of quantum gate sequences. The first cache is a circular buffer
centered around the current instruction address that supports the notion that
the most likely direction is forward motion through the instructions, with
potential local jumps to recently played addresses when looping. The controller
greedily prefetches additional cache lines ahead of the current address but
leaves a buffer of previously played cache lines for looping. Function calls, or
subroutines, require random access so the second instruction cache is fully
associative. The associative cache lines are filled in round-robin fashion
with explicit \texttt{PREFETCH} instructions. This first-in-first-out
replacement strategy for the associative cache ignores any information about
cache line usage. Since the cache controller tracks access, a simple extension
would be a Least Recently Used (LRU) or pseudo-LRU algorithm. It also places a
significant burden on the compiler to insert the \texttt{PREFETCH} instructions
and group subroutines into cache lines. However, given the severe penalty of a
cache miss it is difficult to envisage a hardware-implemented cache controller
that can alleviate that burden.

\subsubsection{Waveform Cache}
\label{subs:waveform-cache}

In use cases we have examined, waveform access does not have the nearly linear
structure of sequence instructions. Rather, a sequence tends to require random
access to a small library of short waveforms, where that library may change over
time due to calibration or feedback signals, or the desire to scan a range of
waveforms. The APS2 has a waveform cache of 128 ksamples to support fast access
to a large waveform library. For scenarios demanding that the library change
over time, the cache is split into two pages of 64 ksamples---see
Fig.~\ref{fig:cache-architectures}(c). The cache is composed of dual-port block
RAM and so a sequence can be actively playing waveforms from one page while the
second page is filled from SDRAM. The two pages' roles can then alternate
supporting total waveform library sizes up to the limit of the SDRAM. For this
mode of operation we do not expect to change the waveform library within a
single sequence. Filling an entire waveform cache page takes $\sim180\units{\mu
s}$, meaning that at typical repetition rates of 10s of kHz we can exchange the
waveform library every few sequences.

\section{Synthesizing and Distributing Steering Information}
\label{sec:TDM}

\begin{figure}
\includegraphics[width=0.95\columnwidth]{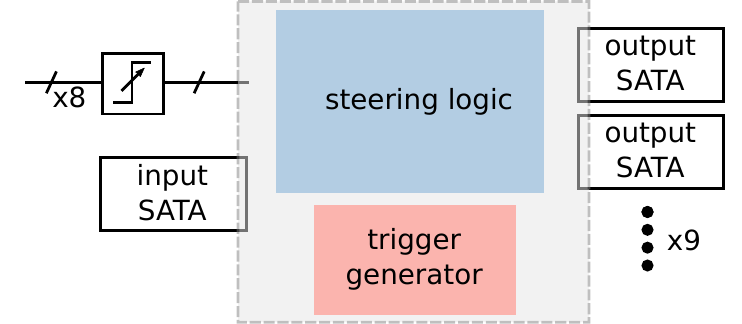}
\caption{ \label{fig:tdm-blocks} Block diagram of the Trigger Distribution
Module (TDM) functionality. 8 SMA inputs to programmable comparators send qubit
state information to the control logic. High-speed serial connections over SATA
cables provide input from other TDM modules and output to APS2 and TDM modules.
The TDM can send a system-wide trigger for intra-crate synchronization}
\end{figure}

As we move beyond simple single qubit feedback circuits we need to synthesize
steering decisions from multiple qubit measurement results, and then communicate
the steering decision to multiple sequencers. We have designed a dedicated
hardware module, the Trigger Distribution Module (TDM), to take in up to eight qubit
state decisions and send steering information to up to nine pulse
sequencers---see Fig.~\ref{fig:tdm-blocks} for a block diagram.

There are eight SMA inputs that feed variable comparators for reading in qubit
measurement results from \texttt{QDSP}, with one input used as a data valid
strobe. The TDM can communicate to all the APS2 modules in an enclosure via a
high speed serial connection over SATA cables. The star distribution network
also allows us to use the distribution module for synchronization. A reserved
symbol acts as a trigger that can be broadcast to all APS2 modules in an
enclosure for synchronous multi-module output. There is one additional SATA
serial link that can be used for inter-crate communications with other TDM's for
for future larger circuits that cannot be controlled with a single crate.

The baseline TDM gateware \texttt{APS2-TDM} (\url{github.com/BBN-Q/APS2-TDM})
currently broadcasts the measurement results to all APS2 modules. As a result,
every APS2 must allow a sequence branch for each result, even when the
controlled qubit is not affected by that particular measurement. A more flexible
decision logic and sequence steering will become critical in larger circuits.
Since all measurement results flow through the TDM, it is natural to consider it
orchestrating the entire experiment. For example, in error correction, syndrome
decoding could be implemented by the TDM and the required qubit corrections sent
to the relevant APS2s only. We see the TDM as a testbed for building out a more
scalable qubit control platform with a hierarchy of controllers, where the TDM
assumes the role of routing measurement results and steering the computation.

\section{Latency}

With all the pieces in place we can examine the latency budget of a closed
feedback loop and highlight potential areas for improvement.  A detailed listing
is provided in Table~\ref{table:latency}. The total latency from the end of a
measurement pulse to the next conditional pulse coming out of the APS2 is
$\approx430\units{ns}$. Our test setup incurs an additional
$\approx110\units{ns}$ of latency from cabling to/from the qubit device in the
dilution refrigerator, as well as analog filtering. The total latency is
comparable to 1\% of the qubit relaxation time and our measurement time, and is
not the limiting factor in our circuit implementation fidelities.

\begin{table}[htb]
  \begin{ruledtabular}
    \begin{tabular}{ l | l }
      Step & Latency (ns) \\
      \hline
      ADC capture                               & 32  \\
      digital signal processing                 & 56 (14 clocks)  \\
      X6 to TDM interface                       & 10 (1 clock)   \\
      TDM distribution logic                    & 10 (1 clock) \\
      TDM to APS2 module interface              & 210    \\
      APS2 address jump                         & 53 (16 clocks)    \\
      APS2 waveform signal processing           & 30 (9 clocks)   \\
      DAC output                                & 29    \\
      \hline
      Total                                     & 428 \\
    \end{tabular}
  \end{ruledtabular}
  \caption{\label{table:latency}Latency budget for closed loop qubit control.}
\end{table}

However, there are a few areas amenable to improvement. The APS2 design
prioritized instruction throughput and waveform cache size. This required
significant buffering and pipelining. Optimizing instead for latency could
tradeoff those capabilities for reduced latency for an APS2 address jump. The
serial link between the TDM and APS2 is slow due to FIFOs that manage data
transfer through asynchronous clock domains. However, synchronizing the TDM and
APS2 to a common $10\units{MHz}$ reference creates a stable phase relationship
between clocks domains which would allow these FIFOs to be removed and save
$\approx 100\units{ns}$. Modest benefit could be obtained by integrating the
readout system into the TDM, saving two data transfer steps.

While not listed in the table, the delays from cabling and analog filtering are
also non-negligible. Since we digitize data at 1GS/s, minimal analog low pass
filtering after mixing down to the IF is necessary, except to prevent
overloading amplifiers or the ADC. Moving the hardware physically closer to the
top of the dilution refrigerator would save $\approx20\units{ns}$. The reduction
in cable delays is one potential benefit to cyrogenic control systems, but is
only a fraction of the total latency budget.

\section{Feedback and feedforward in circuit QED}

The integration of \texttt{QDP} systems and APS2/TDM modules into a circuit QED
apparatus enables a variety of qubit experiments requiring feedback or
feedforward. \emph{Feedback} indicates that measurements modify control of the
measured qubit, while in \emph{feedforward} the conditional control acts on
different qubits. Here we present some examples of simultaneous dynamic control
of up to three qubits. We emphasize that the hardware system was designed for
flexible multi-qubit experiments that allows for programming different
experiments in software, with minimal or no hardware changes.

The quantum processor used here, first introduced in
Ref.~\onlinecite{Riste2017}, is a five-qubit superconducting device housed in a
dilution refrigerator at $\approx10\units{mK}$. The wiring inside the refrigerator is
very similar to the reference above, with the exception of the addition of a
Josephson parametric amplifier (JPA)~\cite{Hatridge2011} to boost the readout
fidelity of one qubit. The control flow of qubit instructions, previously a
pre-orchestrated sequence of gates and measurements, is now steered in real time
by a TDM. This module receives the digital qubit measurements from \texttt{QDSP} digital
outputs, and distributes the relevant data to the APS2 units which then
conditionally execute sequences.

\subsection{Fast qubit initialization}

As a first test of our control hardware, we start with the simplest closed-loop
feedback scheme --- fast qubit reset~\cite{Riste2012, CampagneIbarcq2013}. A
reliable way to initialize qubit registers is one of the prerequisites for
quantum computation~\cite{DiVincenzo2000}. Conventionally, initialization of
superconducting qubits is accomplished by passive thermalization of the qubit to
the near zero-temperature environment. However, with a characteristic relaxation
time $T_1 = 40\units{\mu s}$ (see Table~\ref{table:latency} for relaxation time
details), the necessary waiting constitutes the majority of the experiment wall
clock time. Furthermore, passive initialization slows re-use of ancilla qubits
during a computation, a feature that would relieve the need for a continuous
stream of fresh qubits in a fault-tolerant system~\cite{BenOr2013}.

\begin{figure}
\includegraphics[width=\columnwidth]{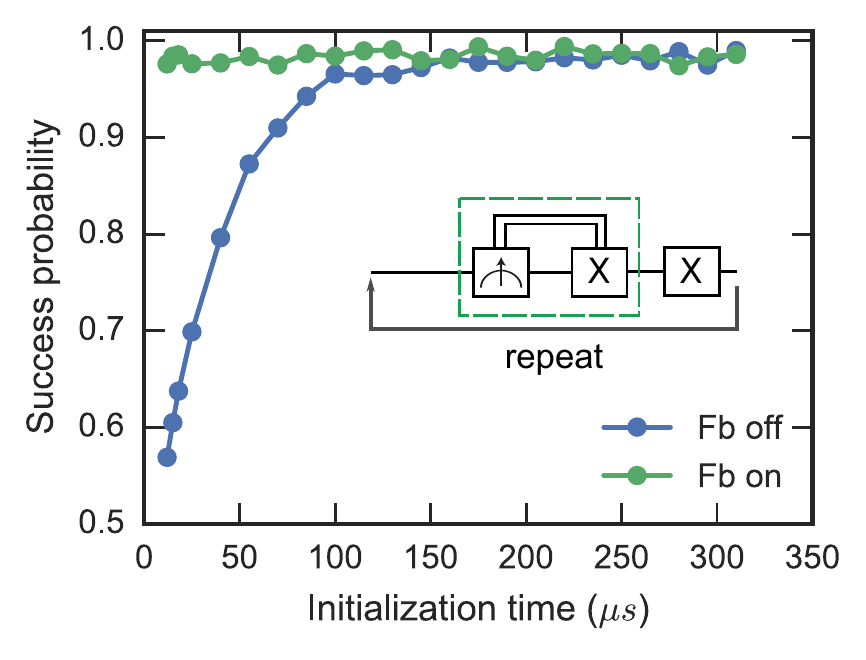}
\caption{ \label{fig:fast-reset} Fast qubit initialization. A simple experiment
consisting of a single $X$ gate is repeated with variable initialization time.
Feedback (green) is used to reset the qubit in the ground state $\ket{0}$ faster
than by waiting for its thermal relaxation (blue). The success probability is
defined as the probability to find the qubit in $\ket{0}$ at the end of each
cycle. Inset: gate sequence per cycle, with a dashed box indicating the feedback
loop. Similar to Ref.~\onlinecite{Riste2012}.}
\end{figure}

Feedback-based reset aims to remove entropy on demand using measurement and a
conditional bit-flip gate (Fig.~\ref{fig:fast-reset} inset)~\cite{Riste2012}.
This operation ideally resets the qubit state to $\ket{0}$ if the measurement
result is $1$, or leaves it unchanged if $0$, giving an unconditional output
state $\ket{0}$. The effect of reset is evident when considering the
initialization success probability compared to no reset (passive initialization)
(Fig.~\ref{fig:fast-reset}). As the initialization time is decreased to $T_1$ or
lower, passive initialization becomes increasingly faulty, while active reset is
largely unaffected.

\begin{figure}
\includegraphics[width=\columnwidth]{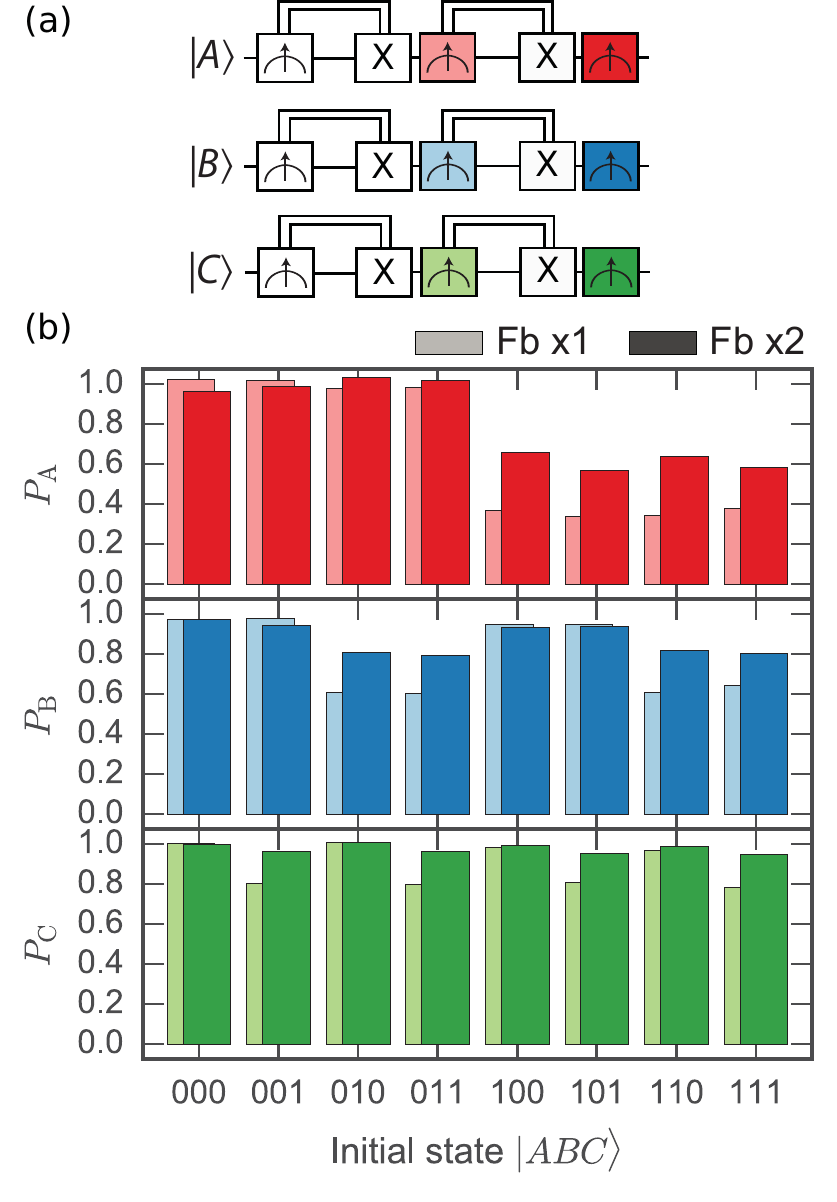}
\caption{ \label{fig:multi-reset} Simultaneous reset of three qubits. (a)
Frequency multiplexed signals are used to independently measure three qubits in
a single \texttt{QDSP} card. A second round of reset can be concatenated to improve
performance. (b) Success probability to reset each qubit measured after 1 (light
bars) and two (dark) rounds. Only one of the readout lines (qubit $C$) is
equipped with a superconducting parametric amplifier~\cite{Hatridge2011},
granting higher readout and reset fidelities.}
\end{figure}

We extend this protocol to reset a register of three qubits simultaneously. This
is accomplished with no additional hardware beyond that already required for the
open-loop control of the same number of qubits. We exploit frequency
multiplexing to combine two readout signals, so that all signal processing can
be accomplished with the two analog inputs of a single X6-1000M. The control
flow simply replicates the conditional bit-flip logic across the three qubits
$\ket{A}, \ket{B}, \ket{C}$ (Fig.~\ref{fig:multi-reset}a). We assess the
performance of the three-qubit reset by measuring the success probabilities for
resetting each individual qubit starting from the eight computational input states
(Fig.~\ref{fig:multi-reset}b). The deviation in success probabilies is largely due
to the difference in readout fidelities (Table~\ref{table:qubits}), as only
qubit $\ket{C}$ is equipped with a JPA.

\subsection{Measurement-based $S$ and $T$ gates}

Our hardware is also readily applicable to feedforward scenarios, where the
result of a measurement conditions the control of different qubits. A first
example is the realization of measurement-based gates. In an error-corrected
circuit, gates on a logical qubit can be made fault-tolerant by applying them
transversally to all the underlying physical qubits. However, for any given
code, a universal gate set cannot all be implemented
transversally~\cite{Xie2008}. For instance, in the surface code, all Pauli
operations $X$, $Y$, $Z$ are transversal, but partial rotations such as
$Z(\pi/2)$ are not.  To fill this gap, fault-tolerant gates can be constructed
with interactions with ancilla qubits and control conditioned on measurement
results~\cite{Bravyi2005}.

\begin{figure}
\includegraphics[width=\columnwidth]{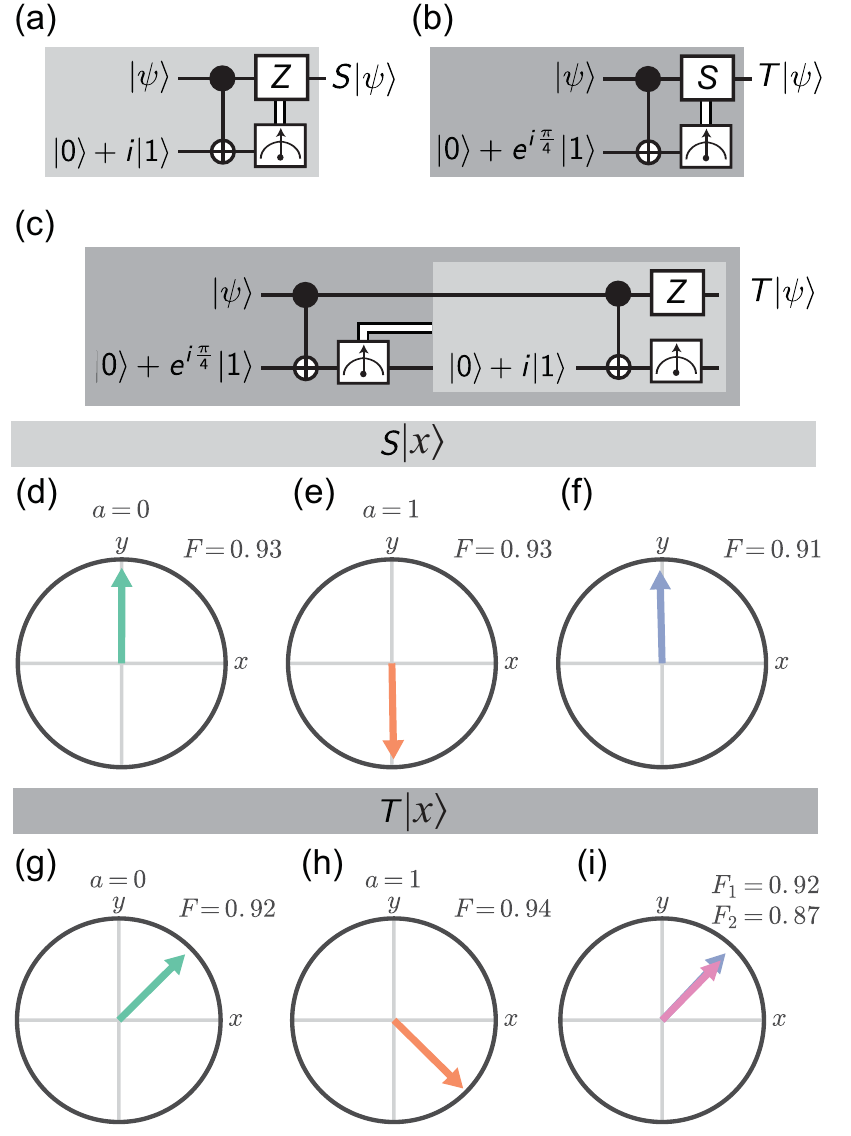}
\caption{\label{fig:Zgates} Measurement-based $S$ and $T$ gates. Gate sequence
to implement $S$ (a) and $T$ (b, c) gates with an ancilla and feedforward. To
construct (c), we replace the $S$ gate in (b) with the circuit from (a). (d-i)
Projected state tomography on the $x$-$y$ plane for initial state $\ket{x} =
(\ket{0} + \ket{1})/\sqrt{2}$ and applied $S$ (d-f) and $T$ (g-i). The data are
postselected on the ancilla measurement result $a=0$ (d, g), $a=1$ (e, h), or
not postselected when feedforward is activated (f,i). The $T$ gate can be made
fully transversal by conditionally implementing the $S$ correction as another
feedforward subroutine. (c, and pink arrow in i).}
\end{figure}

Here we demonstrate the basic principle of measurement-based gates, implementing
partial $Z$ rotations on a physical qubit, using an ancilla and feedforward
operations. The initial state of the ancilla, which can be prepared offline to
the computation, determines the rotation angle $\theta$. Typical gates are
denoted with $S$ ($\theta = \pi/4$) and $T$ ($\theta = \pi/8$). An $S$ gate can
be decomposed into an ancilla measurement and a conditional $Z(\pi)$
gate~\cite{Bravyi2005}, which is transversal in the surface code
(Fig.~\ref{fig:Zgates}a). Starting with the ancilla in a superposition state,
$\ket{\psiz} = (\ket{0} + \ket{1})/\sqrt{2}$, the result of the ancilla
measurement determines whether the final state approximates the desired
$S\ket{\psiz} = (\ket{0} + i\ket{1})/\sqrt{2}$ (Fig.~\ref{fig:Zgates}d), or the
$\pi$ shifted $ZS\ket{\psiz}$ (e). In the latter case, a corrective $Z$, applied
as a frame update (see Sec.~\ref{subs:modulation-engine}), gives the intended
state $S\ket{\psiz}$  deterministically (f). The reduced coherence, indicated by
the length of the arrow, is mainly due to the measurement time ($0.9\units{\mu s}$), with
the addition of $\sim 0.54\units{\mu s}$ decision latency in (f).

Similarly, a $T$ gate can be implemented with a different ancilla preparation
and a conditional $S$ gate (Fig.~\ref{fig:Zgates}b). However, as seen before,
the $S$ gate cannot be applied transversally, so it is in turn decomposed into
the feedforward sequence above. The result is a nested feedforward loop with up
to two ancilla measurements and conditional sequences (Fig.~\ref{fig:Zgates}c).
We reuse the same ancilla in the second round, taking advantage of the first
measurement to initialize it in a known state. By using the CLEAR
protocol~\cite{McClure2016}, we reduce the latency before we can reuse the
ancilla (Fig.~\ref{fig:Zgates}g-i).

\subsection{Entanglement generation through measurement}

\begin{figure}
\includegraphics[width=\columnwidth]{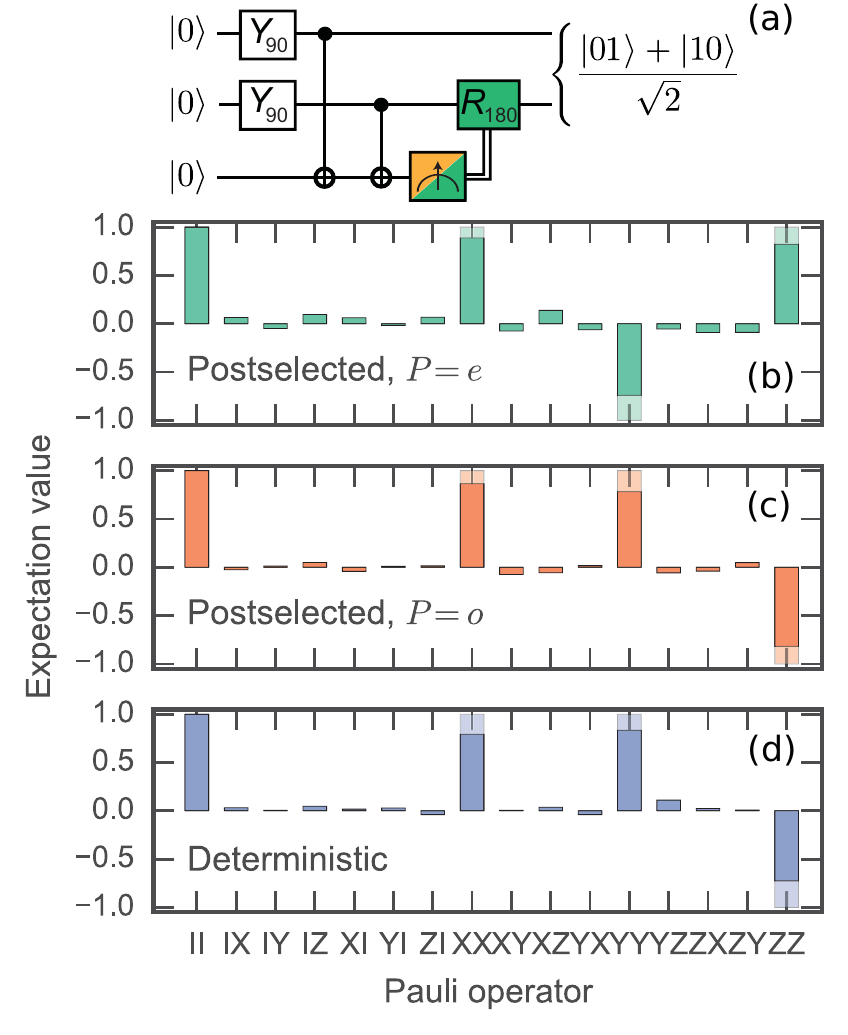}
\caption{ \label{fig:ebm} State tomography of entanglement by measurement and
feedforward. The post-selected result of a two-qubit parity measurement (a)
determines whether the qubits are projected onto an even (b) or odd (c)
entangled state~\cite{Chow14, Saira14}. Programming feedforward control that
conditionally switches the parity from even to odd generates the target
entangled state deterministically~\cite{Riste2013} (d). State tomograms shown in
the Pauli basis. Opaque (transparent) bars indicate the measured (ideal)
expectation values for the two-qubit Pauli operators.}
\end{figure}

With three qubits, feedforward control can be used to generate entanglement by
measurement. Two qubits separately interact with a third ancilla qubit to
implement a parity measurement of the first two qubits (Fig.~\ref{fig:ebm}a).
With the first two qubits starting in an equal superposition state, the parity
measurement projects them onto either an even or odd Bell state with the ancilla
measurement result containing the information about which
(Fig.~\ref{fig:ebm}b-c). This parity measurement scenario, with ancillas and
feedforward, is also relevant for syndrome extraction in quantum error
correction schemes~\cite{Bravyi1998, Mermin2007} and has been experimentally
demonstrated in post-selected form\cite{Saira14, Chow14}. With our hardware we
can go one step further, and deterministically create the odd state by
converting the projected even state into an odd one by a conditional bit-flip on
one of the data qubits (Fig.~\ref{fig:ebm}d). This deterministic protocol has
also been realized in Ref.~\onlinecite{Riste2013}, but with the ancilla qubit
replaced by a cavity mode.

\begin{table*}
\label{table:qubits}
  \begin{ruledtabular}
    \begin{tabular}{ l | l | l | l }
        & {Fig.~\ref{fig:fast-reset}} & Fig.~\ref{fig:multi-reset} & Fig.~\ref{fig:ebm} \\
      \hline
      Measurement time ($\mu$s)  &     2.2     &  4.5    &     2.2 \\
      Characteristic relaxation time $T_1$ ($\mu$s) & $\sim20$ &  $\sim40$ ($A$), 40 ($B$), 20 ($C$)  &  $\sim20$\\
      Measurement assignment fidelity  & 0.95 &  0.70 ($A$), 0.81 ($B$), 0.94 ($C$)  &  0.95 \\
    \end{tabular}
  \end{ruledtabular}
  \caption{Relevant measurement and qubit parameters for the experiments in Figs.~\ref{fig:fast-reset}-\ref{fig:ebm}.}
\end{table*}

\section{Conclusion}

The APS2 and \texttt{QDSP} platforms are a complete hardware solution for
dynamic quantum computing systems. They achieve this with tailored gateware and
hardware that enable flexible, low-latency manipulation, thus allowing users to
program generic quantum circuits without hardware reconfiguration. We have
proved this hardware \emph{in situ} with a superconducting quantum processor,
showing a variety of novel dynamic circuits utilizing feedback and feedforward.
To further improve this platform we intend to integrate control and readout into
a unified hardware system, investigate improvements to the APS2 analog output
chain and generalize system synchronization.

Upconversion systems generically require a multitude of components and suffer
from various mixer imperfections, leading to instability and a spectrum polluted
by mixer product spurs. Future hardware revisions may solve these issues by
moving to faster RF DACs that can directly generate microwave tones with a
cleaner spectrum~\cite{Glascott-Jones2014}. Direct RF output allows for greater
frequency agility, allowing for channel re-use for both control and measurement.
New DACs with sampling rates from 4--6$\units{GS/s}$ support output modes that
direct power into higher Nyquist zones, removing pressure for ultra-high clock
speeds. Future FPGAs may include many on-chip RF DACs~\cite{Erdmann2017},
potentially drastically increasing channel densities in control systems.

The typical way to achieve system synchronization is by building trigger fanout
trees. This strategy becomes increasingly cumbersome and fragile as system sizes
grow. A more scalable approach consists of sharing frequency and time between
all devices, so that all modules in the system have a synchronous copy of a
global counter. To achieve this, future hardware revisions may incorporate a
time distribution protocol such as White Rabbit~\cite{Serrano2009}. Sharing time
changes the synchronization paradigm from ``go on trigger'' to ``go at time $t$''.

Finally, we are exploring methods to combine real-time computation with dynamic
control-flow on the individual APSs. For example, a controller of a system of
logical qubits must combine information from a logical decoder with program
control-flow. A softcore CPU running on the TDM would enable rapid development
of realtime infrastructure.

\begin{acknowledgments}

Schematic capture and PC board layout for the APS2 and TDM were done by Ray
Zeller and Chris Johnson of \emph{ZRL Inc.}, Bristol, RI. Nick Materise
developed an initial prototype of the \texttt{QDSP} system in VHDL. This was
converted into a Simulink model and tested with MATLAB HDL Coder before finally
being converted back into pure VHDL. The data analysis for the experimental
section was performed using code written in Julia~\cite{Bezanson2014}, and the
figures were made with Seaborn~\cite{seaborn_v0.7.1} and
matplotlib~\cite{Hunter2007}. We used Scipy~\cite{scipy} to construct the filter
coefficients for the \texttt{QDSP} system. The authors would like to thank
George A. Keefe and Mary B. Rothwell for device fabrication, and Nissim Ofek for
discussions about AWG instruction sets. This research was funded by the Office
of the Director of National Intelligence (ODNI), Intelligence Advanced Research
Projects Activity (IARPA), through the Army Research Office contract
No.~W911NF-10-1-0324 and No.~W911NF-14-1-0114. All statements of fact, opinion
or conclusions contained herein are those of the authors and should not be
construed as representing the official views or policies of IARPA, the ODNI, or
the U.S. Government.

\end{acknowledgments}

\bibliography{references}

\end{document}